\documentclass[twocolumn,amsmath,amssymb]{revtex4}

\usepackage{graphicx}
\usepackage{dcolumn}
\usepackage{bm}

\begin{document}

\title{Controlling the direction and amount of a splash with textured surface}
\author{Lei Xu}
\author{Sidney R. Nagel}
\affiliation{The James Franck Institute and Department of Physics\\
The University of Chicago\\ 929 East 57th St., Chicago, Illinois
60637}

\date{\today}

\begin{abstract}
Because splashing is such a violent process, one might naively
expect that neither the direction of droplet emission nor the amount
of ejected material can be controlled with any precision.  Even
though it is observed countless times in the course of a day, drop
splashing is highly non-intuitive\cite{Quere1}: for example, it is
surprising that the surrounding air pressure determines whether a
drop will splash when hitting a smooth surface\cite{Xu1}.  Here we
describe a discovery with significant practical ramifications: the
direction, as well as the number, of the ejected droplets can be
controlled by the texture of a surface.
\end{abstract}
\maketitle

    Splashing of a liquid drop after hitting a solid surface is crucial
in many industrial applications such as inkjet
printing\cite{inkjet}, surface coating\cite{coating}, combustion of
liquid fuel\cite{combustion} and spray drying\cite{drying}. Precise
manipulation of the direction in which ejected droplets emerge from
the point of impact as well as of the number and total volume of the
ejecta is highly desirable. We find that all of these parameters can
be controlled by using a substrate with a well-defined texture.

    Using UV-lithography techniques, we create a substrate consisting
of a square lattice of square pillars as shown in Fig. 1a.  The
three parameters, (i) vertical pillar height, $h$, (ii) lateral
pillar side, $l$, and (iii) spacing between pillars, $s$, can be
varied independently.  Fig.1b shows bottom-view photographs of a
$3.4 mm$ diameter drop after colliding with this substrate at $4.3
m/s$ in a low-pressure environment of $13 kPa$.  The liquid is a
mixture of ethanol with ink so that the splash is clearly visible.
There is a striking feature: the splashes have 4-fold-symmetry. They
emerge predominantly in the diagonal directions of the underlying
lattice where the distance between pillars is the largest.  This is
consistent with data\cite{Xu2} showing that in this regime of pillar
dimensions, splashing is suppressed by a higher pillar density.

\begin{figure}[!h]
\begin{center}
\includegraphics[width=3.2in]{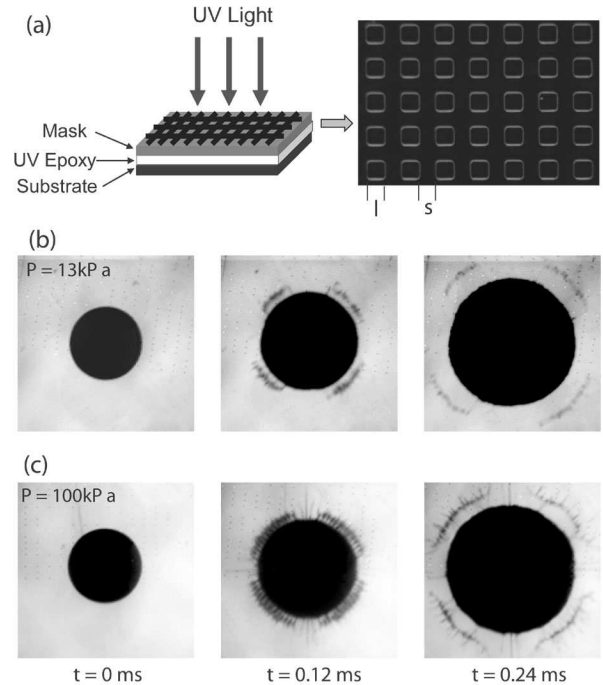}%
\caption{ Splash on a textured surface. (a), Textured surface
manufactured by UV-lithography.  The left cartoon shows the
lithography process.  The photograph at right is a top view of the
textured surface.  Each square is a pillar with height $h = 18\mu
m$; the lateral pillar size, $l$, is the same as the pillar spacing,
$s$: $l = s = 60\mu m$. (b), A splash at low pressure, $13kPa$.
There is a clear 4-fold-symmetry, predominantly in diagonal
directions of the lattice.  (c), A splash at atmospheric pressure,
$100kPa$. The amount of splash increases with air pressure, but the
4-fold-symmetry is preserved.}
\end{center}
\end{figure}

    There are two distinct components to a splash: the corona splash
caused by air\cite{Xu1} and the prompt splash caused by surface
roughness\cite{Xu3}. By working at low pressure, as in Fig. 1b, air
effects are minimized so that there is no corona splash.  In this
case only the surface-roughness induced prompt splash remains. Fig.
1c shows another series of photographs of drop splashing with the
same initial conditions as in Fig. 1b except that the background air
is at atmospheric pressure, $100 kPa$.  Here both the air and the
surface roughness contribute to producing a splash.  We see a larger
splash but with the same 4-fold-symmetry seen in Fig. 1b. This
indicates that even though air is present, the splash directions are
still controlled by the substrate texture.

    The surface texture not only controls the direction of the
droplets ejected by a splash, but it also affects the splash
magnitude.  To show this effect, we did a series of experiments at
atmospheric pressure with fixed lateral pillar dimensions: $l = s =
60\mu m$, while varying the pillar height, $h$. The impact velocity
was kept at $V_0 = 4.3m/s$.  Fig. 2a shows a symmetric corona splash
on a smooth surface ($h = 0\mu m$). The photographs in Fig. 2b show
that for small $h$ ($h = 5\mu m$), there is a large splash.  This
splash is not as symmetric as the smooth case.  Fig. 2c shows that
as we increase the height to $h = 54\mu m$ (a value close to $l$ and
$s$), the splash amount is greatly reduced.  Finally, Fig. 2d shows
that further increasing the height to $h = 125\mu m$ completely
suppresses all splashing.

\begin{figure}[!t]
\begin{center}
\includegraphics[width=3.2in]{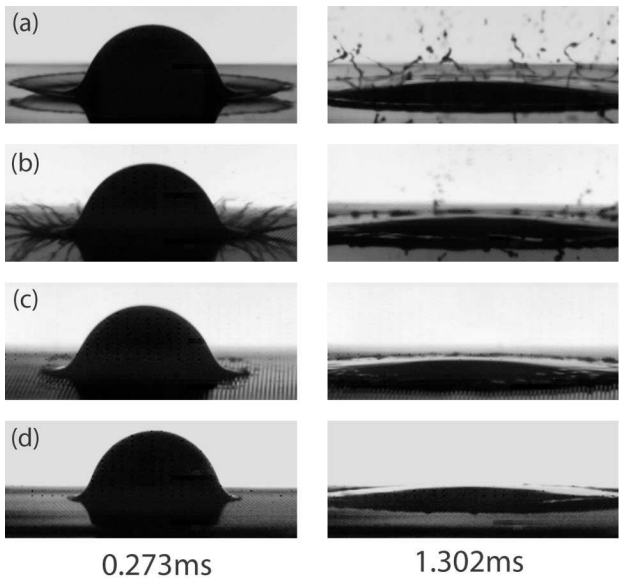}%
\caption{ Splashing on surfaces with different pillar heights under
atmospheric air pressure.  In all images, the lateral pillar
dimensions were fixed at $l = s = 60\mu m$. (a) A symmetric corona
splash on a smooth surface ($h = 0\mu m$). (b) When $h = 5\mu m$,
the splash is a mixture of a corona and a prompt component. The
ejected droplets are not as symmetrically distributed as in (a). (c)
As we increase $h$ to about the same size as $l$ and $s$, $h = 54\mu
m$, there is a dramatic decrease in the magnitude of the splash. (d)
When $h$ is further increased to $h = 125\mu m$, the splash is
completely suppressed.}
\end{center}
\end{figure}

    Why do tall pillars suppress splashing?  Tall pillars can form
channels through which air can easily escape without interfering
with the liquid motion.   This can minimize the effects of air that
would otherwise create a corona splash.  However, it is much less
clear why tall pillars would also reduce the contributions from the
prompt component of the splash.  This result suggests that the drop
feels little disturbance during its expansion on top of the tall
pillar surfaces.  This could resemble the drops floating on top of
the super-hydrophobic surfaces studied by Qu\'{e}r\'{e} et al
\cite{Quere2, Quere3}.

    Thus texture can control both the direction and magnitude of a
splash.   Since splashing is involved in many industrial processes,
these discoveries should have important applications.

We wish to thank Qiti Guo, Jingshi Hu, David Qu\'{e}r\'{e}, Mathilde
Callies-Reyssat, and Wendy Zhang for helpful discussions.  This work
was supported by MRSEC DMR-0213745 and NSF DMR-0352777, L. X. was
supported by Grainger Fellowship.

\end{document}